# From Counting Electrons to Calibrating Ammeters: Improved Methodologies for Traceable Measurements of Small Electric Currents

Speaker: Stephen P. Giblin, National Physical Laboratory, Hampton Road, Teddington, Middx. TW110LW, U.K. Phone: +44 20 8943 7161. Email: Stephen.giblin@npl.co.uk.

Author: Jordan Tompkins, National Physical Laboratory, Hampton Road, Teddington, Middx. TW110LW, U.K. Phone: +44 20 8943 8773. Email: jordan.tompkins@npl.co.uk.

New technology, the ultrastable low-noise current amplifier and the electron pump, provide new methods for making traceable measurements of small DC electric currents. We review four traceability routes for small current measurements and discuss the merits of each one. We present three case studies of small current calibrations, highlighting the role of noise and drifting instrument offsets. We show how the Allan deviation is used as a statistical tool for designing a calibration cycle to correctly eliminate drifting instrument offsets from calibration data. We also present a simplified noise model for a low-current ammeter which predicts a lower limit to the achievable statistical uncertainty in a calibration.

1. Introduction

The traceable measurement of small electric currents is becoming increasingly relevant in a number of sectors. By "small current" we mean currents below 1 nA, where measurement functionality is not available in standard off-the-shelf multimeters, and specialised instrumentation must be deployed. Examples of measurement areas requiring small current traceability are radiation dosimetry, environmental particulate monitoring, semiconductor characterisation and characterisation of electrical insulators. Against these burgeoning needs are set a number of challenges. Although progress has been made in National Metrology Institutes (NMIs) over the last 10-15 years, traceability routes for small currents are still not widely available, and commercial instruments suffer from instabilities which limit the accuracy achievable outside specialised NMIs. With very small signals to measure, obtaining a satisfactory signal-to-noise ratio can require long averaging times, and this in turn requires the correct use of experimental methodology and statistical data analysis to assign a meaningful random uncertainty to the measurement. In this paper, we review the available traceability paths for small currents, including two key new developments, the ULCA [1] and the electron pump [2] which are likely to play an increasingly important role in small-current metrology. We consider the noise properties of typical small-current ammeters and show how these constrain the uncertainty of a calibration at different current levels. Finally, we present three case studies of small-current instrument calibrations, illustrating the use of different traceability routes and highlighting important points of best practice.

## 2. Traceability routes

Four traceability routes for reference currents are summarised schematically in figure 1. Direct electro-mechanical methods of generating a reference current according to the present SI definition of the ampere are no longer used, and most realisations start with the Josephson voltage standard (JVS) and quantum Hall resistance (QHR). We now discuss each of these methods in turn.

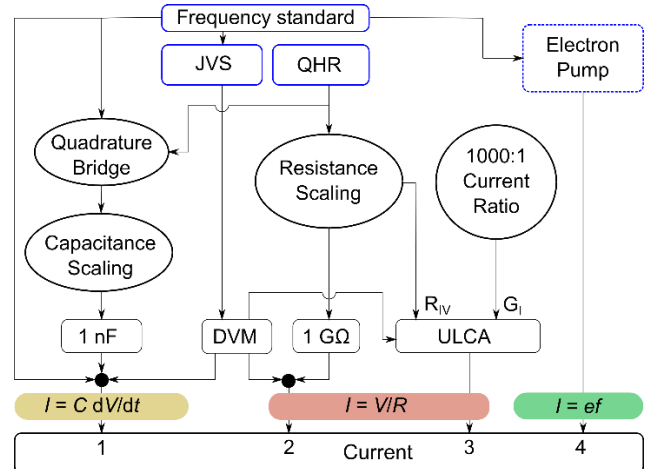

Figure 1. Schematic diagram of 4 traceability routes for small currents. JVS = Josephson Voltage Standard, QHR = Quantum Hall resistance, ULCA = Ultrastable low-noise current amplifier. Primary standards are indicated by blue boxes, with the dotted blue box around electron pump indicating that it is a prototype standard under development. In the formula $I=ef$, $e$ is the charge on the electron.

### 2.1 The capacitor ramp method

Applying a linear voltage ramp to a low-loss capacitor (method 1 in Fig. 1) is probably the most widely used method of generating a sub-nA reference current, and several systems have been reported with uncertainties of the order of tens of parts per million (ppm) [4-7]. The reverse process, of measuring a current by using an integrator, has been a common method of small current measurement for many decades, particularly in ionising radiation metrology where the activity of a radioactive source is proportional to the current output from an ionisation chamber [8]. From a traceability point of view, it is easy to see why this method is so popular: The three parameters capacitance, voltage and time can all be readily obtained as standard off-the shelf calibrations. For currents below 1 nA and voltage ramp rates below 1 V/s, the required values of capacitance fall in the range 1 pF to 1 nF, where air or sealed-gas dielectric capacitors are commercially available, and audio-frequency bridge techniques exist for calibrating these capacitors with uncertainties around $10^{-6}$. Precision DVMs such as the Keysight 3458A or Fluke 8508A[1] have uncertainties around the $10^{-6}$ level for long intervals after calibration, and simple quartz timing circuits can likewise achieve $10^{-6}$ uncertainty.

The crucial un-resolved issue with capacitor ramp current sources concerns possible frequency dependence of the capacitors. This is relevant because the capacitor is calibrated at audio frequency, typically 1 kHz, but used in the ramp generator at a frequency many orders of magnitude lower, typically 10 mHz. One study investigated the frequency dependence of a sample of sealed-nitrogen and air-dielectric standard capacitors of 100 pF and 1 nF value over this frequency range, and found worst-case relative corrections of nearly $2\times10^{-4}$ for one example of a sealed-nitrogen capacitor and $3\times10^{-4}$ for an unsealed air-dielectric capacitor [9]. These corrections are large compared to the other uncertainties associated with the quantities $C$(1 kHz), $V$ and $t$. Determining the frequency-dependence correction is a laborious process [9], and following this finding, the author's laboratory stopped using the capacitor ramp method as a primary method for generating current. We focussed instead on resistor-based current sources, discussed in the next subsection.

---
[1] Here and throughout this paper, specific commercial instruments are mentioned for information only and not for endorsement purposes.

### 2.2 Resistor and voltage source

The generation of a current by applying a voltage across a resistor (method 2) is limited in accuracy, for small currents, by the increase in the calibration uncertainty with increasing resistor value. At the National Physical Laboratory (NPL), decade resistors up to 1 GΩ are calibrated using a cryogenic current comparator (CCC), which replaces Hamon scaling techniques used previously. The calibration and measurement capability (CMC) uncertainty for calibrating 1 GΩ resistors at NPL is $0.8\times10^{-6}$, and a re-evaluation of the CCC uncertainty, currently underway, yields a provisional revised uncertainty of around $0.2\times10^{-6}$. In the NPL reference small current generator (inset to figure 3a), a 1 GΩ standard resistor is supplied with a voltage monitored using a precision DVM calibrated regularly against a JVS, and total relative uncertainties in the output current of around $0.3\times10^{-6}$ can be achieved. The only assumption is that the resistor has negligible power coefficient, since it is calibrated with currents from 10 nA to 100 nA, and used in the reference current source with currents less than 1 nA. Since this current range corresponds to very small power dissipation, 1 nW to 10 µW, we believe this assumption is reasonable, and it is supported by measurements on similar standard resistors [10] Two minor disadvantages of this technique compared to the capacitor ramp method are worth noting. The first concerns the output resistance of the current source; calibrating an ammeter with an unusually high input resistance may require a correction for the input resistance of the ammeter, depending on the desired uncertainty level. The second is that the 1 GΩ resistor adds thermal noise to the reference current, in contrast to a capacitor-based source. This small extra noise, discussed quantitatively in section 3, may increase the statistical uncertainty for calibrations of the most sensitive ammeters. These two compromises are worth making in most cases, in return for the lower type B uncertainty offered by this route. This reference current source has been used very successfully to evaluate the performance of electron pumps – prototype current standards which operate by controlled transport of single electrons. These devices are discussed in section 2.4.

### 2.3 The Ultrastable low-noise current amplifier (ULCA)

The ULCA (method 3) is a novel solution to the problem of ultrahigh-accuracy small current sourcing and measuring developed at the Physikalisch-Technische Bundesanstalt (PTB), Germany [1,11-13]. In measurement mode, the standard ULCA functions as a current-to-voltage converter with nominal gain $10^9$ V/A, and the voltage output is measured using a DVM traceable to the JVS. The ULCA can also be configured to operate as a current source. Very recently, new variants of the ULCA with nominal gains ranging from $10^8$ V/A to $10^{12}$ V/A have been developed [13] although we refer just to the standard ULCA in this paper. The current-to-voltage gain is obtained from two functional blocks internal to the instrument which must be calibrated separately: a 1000:1 current gain stage, and a $10^6$ V/A transimpedance gain stage calibrated as a 1 MΩ resistor.

Although these two calibrations require sophisticated metrology infrastructure (ideally, a CCC), the ULCA gain has exhibited remarkable stability – around 1 part in $10^6$ - following international transport by commercial courier [12], and annual drift at a fixed location of a few ppm [1]. These numbers are roughly two orders of magnitude smaller than can be typically expected from commercial low-current ammeters. The implication is that an ULCA calibrated at an NMI could be used as a reference current source (or indeed as an ammeter) at another laboratory with a type B uncertainty less than 10 ppm, which is sufficient for most industrial calibration purposes. It can therefore be envisaged that the ULCA will play a major role in small current traceability both at NMIs and commercial calibration labs in the future.

2.4 Electron pumps

The idea of making a current standard by controlling electron flow one electron at a time dates back nearly 30 years, and the first experiments with metrological accuracy were performed at NIST, Boulder, in the mid- to late-1990s [14]. Conceptually, the idea is very elegant. Using micro- and nano- fabrication techniques it is possible to make devices which transport one electron for each cycle of a periodic control voltage at frequency $f$, generating a current, in the absence of errors, equal to $ef$ where $e$ is the charge on the electron. These devices are called 'electron pumps', by analogy to water pumps, for the technical reason that they can pump electrons 'uphill', against a small bias voltage, although for metrological applications they are operated at close to zero bias voltage. As shown in figure 1 (method 4), electron pumps generate a reference current with an extremely straightforward traceability route, requiring only a known frequency input. Unlike the previous three routes, no scaling devices or other metrological infrastructure is required. Making electron pumps with a combination of high accuracy and high output current has proved extremely challenging, however. The early NIST devices were limited to currents of $\approx 1$ pA. Prototype capacitance standards, based on measuring the voltage across a capacitor charged with a known number of electrons, were subsequently realised at both NIST [15] and PTB [16]. In the meantime, semiconductor tunable-barrier pumps were developed [17], several designs of which have demonstrated currents as high as 160 pA with a relative uncertainty of $10^{-6}$ or less [18-22]. At the present time, the bench-mark is a current of 91 pA equal to $ef$ to within a measurement uncertainty of $1.6\times10^{-7}$ [21]. One recent experiment on an electron pump based on a single atomic trap in silicon demonstrated pumping at the remarkably high current of 1.2 nA, albeit with a relative error of $2\times10^{-5}$ [23]. The rapid progress made in recent years suggests that practical calibrations of nano-ammeters using electron pumps are within reach. The very simple traceability route, combined with the fact that under the future revised SI system an electron pump will directly realise the unit ampere, makes them very attractive as primary current standards, and they are currently the focus of research at several NMIs.

3. General noise considerations

Before we consider details of specific cases of low-current ammeter calibrations, we will develop an approximate theory predicting the statistical uncertainty of a current measurement with a given averaging time. This is possible because although commercially-available low-current ammeters differ in detail, the majority of them are nowadays designed as transimpedance amplifiers. In this type of circuit, an amplifier with low input current noise uses a high-ohmic feedback resistor $R_F$ to convert the input current $I_{IN}$ to an output voltage $V_{OUT} = I_{IN}R_F$ (upper right inset to Figure 2). If the ammeter is a stand-alone unit, $V_{OUT}$ is digitised by an analogue-to-digital converter internal to the instrument and converted to a current reading in firmware. Stand-alone transimpedance amplifiers are also available, which require the user to connect their own voltmeter to measure $V_{OUT}$. Transimpedance amplifiers generally have a complex noise response as a function of frequency [3], but at low frequencies, less than about 10 Hz, we can derive a simplified noise model and draw some general conclusions about the achievable random uncertainty for a given averaging time.

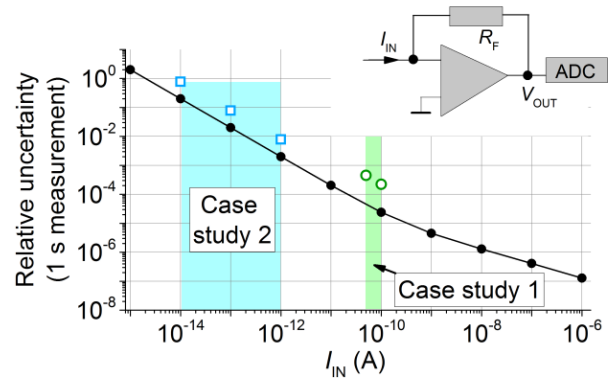

Figure 2. Solid black points and connecting lines: relative statistical uncertainty for a 1 second measurement time, as a function of input current for the simplified ammeter noise model described in the text. Open points show the statistical uncertainties obtained in calibrations discussed in sections 4.1 (circles) and 4.2 (squares) and the coloured rectangles demarcate the current range covered by those case studies. The upper right inset is the circuit of a Transimpedance amplifier

In our simplified noise model, we consider just two contributions to the total noise. The first is the intrinsic input current noise of the amplifier $I_N$, which for a well-designed amplifier should be less than 1 fA/√Hz. The second contribution is due to thermal noise (also known as Johnson-Nyquist noise) in the feedback resistor $R_F$. In a 1 Hz bandwidth, this is given by $I_R = \sqrt{(4k_BT/R_F)}$, where $k_B$ is Boltzmann's constant and $T$ is temperature. For $R_F = 1$ GΩ at room temperature, $I_R = 4.2$ fA/√Hz, and for $R_F = 1$ TΩ (close to the practical upper limit for a feedback resistor), $I_R = 0.13$ fA/√Hz. The total noise is the root-sum-square of the two contributions, $I_T = \sqrt{(I_N^2 + I_R^2)}$, and the relative statistical uncertainty for a 1 s measurement is $I_T/I_{IN}$. In a typical commercial instrument, a number of ranges will be available in which different values of $R_F$ can be selected, and following best practice, the user will select the range which makes the output voltage $V_{OUT}$ close to the full scale of the digitiser. In figure 2, we plot the statistical uncertainty for decade values of input current, assuming $I_N = 0.2$ fA/√Hz and $R_F = 1$ V/$I_{IN}$, appropriate for a digitiser with ± 1V full scale. Two regimes are apparent. For currents larger than about 100 pA, $I_R \gg I_N$, and the relative uncertainty decreases as the square root of the input current. As the current increases further, the relative statistical uncertainty goes below $10^{-5}$, and the type B uncertainty in the reference current source is likely to dominate the uncertainty budget of most ammeter calibrations. In the low-current regime, below about 100 pA, $I_N \gg I_R$, and the relative uncertainty decreases linearly with input current. For most calibrations in this regime, the statistical, or type A, uncertainty in the ammeter readings will dominate the uncertainty budget unless very long measurement times are used. We emphasise that the uncertainty in a real calibration will be higher than in the idealised model presented here, due to a number of factors including possible noise due to the reference current source itself, and additional noise sources in the ammeter not included in our simple noise model. For example, capacitance at the input can produce capacitive noise gain even at low frequencies in sensitive ammeters [3], and for this reason the cable connecting the reference current source to the ammeter should be as short as possible.

The actual relative statistical uncertainties, scaled for a 1 s measurement time, obtained from calibration data discussed in sections 4.1 and 4.2 are plotted as open symbols in Fig. 2. These data were obtained from calibrations of two very different ammeters, and yet they track the theoretical line quite closely. This agreement supports the validity of the approximate noise model presented in this section.

4. Calibration Case studies

In this section, we present three case studies of calibrating low-current measuring instruments with different characteristics and for different applications. We will refer to the instrument being calibrated as the Device Under Test (DUT). First, it is important to define precisely what we mean by "calibration". Generally, the current indicated by the instrument can be expressed as $I_M C = I_{IN} + I_{OFF}$. In this equation, $I_{IN}$ is the unknown current supplied to the instrument's input, $C$ is the calibration factor and $I_{OFF}$ is the offset current – in other words, the current displayed by the DUT when the input current is zero. In the cases considered in this paper, the purpose of the calibration is to determine the calibration factor $C$. The offset current $I_{OFF}$ is usually too strongly dependent on time and environmental parameters to be worthwhile calibrating, and measurement practitioners should have procedures in place for subtracting $I_{OFF}$ from a measured result.

To remove $I_{OFF}$ from the calibration data, during the calibration the DUT is supplied with known input currents at two levels, $I_{IN1}$ and $I_{IN2}$, yielding two indicated values $I_{M1}$ and $I_{M2}$, with $I_{IN2}$ often chosen to be zero, or equal to $-I_{IN1}$. The calibration factor is then given by $C = \Delta I_{IN}/\Delta I_M = (I_{IN1}-I_{IN2})/(I_{M1}-I_{M2})$, under the assumption that $I_{off}$ is constant on the timescale required to make the two measurements.

4.1 Picoammeter for readout of ion chamber current

Here, we calibrate a commercial low-current ammeter, the Keithley 6430, using the NPL reference current system described in section 2.2. The intended application is the readout of current from an ionisation chamber for radionuclide metrology, but this instrument could be used in a range of applications. The target uncertainty is 100 ppm. The calibration is at two indicated current values, 50 pA and 100 pA, on the DUT's 1 nA range. The Keithley 6430 measures current flowing into the instrument to have a negative sign because of its source-measure architecture, and thus, we choose $I_{IN1}$ to be either -50 pA or -100 pA, and $I_{IN2}$ to be zero.

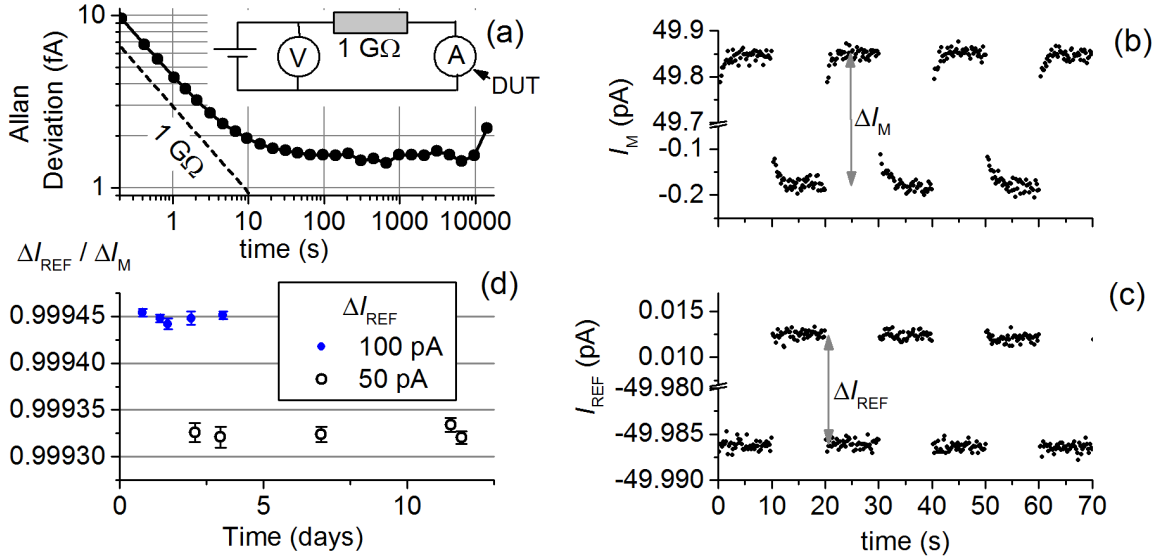

Figure 3. Calibration of a Keithley 6430 at nominal current levels of 50 pA and 100 pA, on the 1 nA range. Panel (a): Allan deviation as a function of measurement time, computed from a time-series of ammeter readings, with the ammeter connected to the NPL reference current source. The theoretical Allan deviation due to the thermal noise in 1 GΩ is shown as a dashed line, and the inset shows a schematic of the circuit. Panel (b): Section of raw ammeter readings from a calibration run at nominal 50 pA current, showing 3 ½ on-off cycles. Panel (c): section of raw reference current source readings from the same run as (b). Panels (b) and (c) share the same time axis, and vertical double arrows indicate the differences $\Delta I_M$ and $\Delta I_{REF}$. Panel (d): Calibration factor from repeat calibrations spread over several days. Each data point is averaged from 100 on-off cycles, and the error bars indicate the statistical uncertainty.

One key consideration concerns the assumption that $I_{OFF}$ is constant on the timescale required to make the two measurements $I_{M1}$ and $I_{M2}$. If $I_{OFF}$ is drifting rapidly, the two measurements must be made in a shorter time interval, and conversely, if $I_{OFF}$ is quite stable, a longer interval could be used. The Allan deviation [24] is an invaluable tool in this context. Originally developed for time and frequency metrology to quantify the stability of frequency standards, it is a type of analysis which can be applied to any measured quantity where the data takes the form of readings spaced evenly in time. It directly yields an estimate of the statistical uncertainty after a given measurement time. In the example at hand, we first acquire a continuous series of about $10^5$ readings from the DUT. Using an integration time per reading of 0.2 s, this takes a couple of hours. Then we apply the Allan deviation analysis to the time-series. We will not describe the analysis in detail here, but it is simple to implement in software, and readers are referred to ref [24] for the formula. Figure 3(a) shows the Allan deviation of the DUT on the 1 nA range, connected to the NPL reference current source set to zero output current. For short averaging times, the Allan deviation decreases as the square root of the averaging time, which is the familiar result expected in the presence of frequency-independent noise (the longer the measurement, the lower the statistical uncertainty). The dashed line in the figure shows the Allan deviation expected for thermal noise in the reference current source. This is the noise level we would expect to measure using an ideal noiseless ammeter. The DUT shows an elevated noise due to its internal feedback resistance. For times much longer than about 10 s, the Allan deviation becomes roughly independent of time due to the drifting offset: *increasing the averaging time does not decrease the statistical uncertainty*. The cross-over point defines the time-scale for the two parts of the calibration. Ideally in this case we should measure $I_{M1}$ and $I_{M2}$ in less than 10 s, but we cannot make the measurement time too short because the DUT has a finite settling time following a change in the input current. We chose a compromise of 20 seconds; 10 seconds measuring $I_{M1}$ and 10 seconds measuring $I_{M2}$.

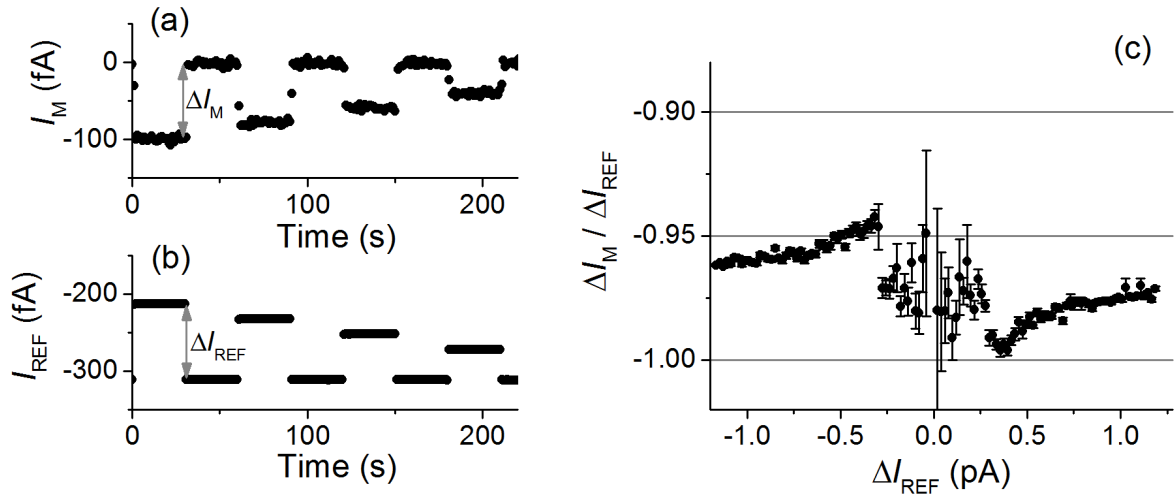

Figure 4. Calibration of a Grimm model 5.705 at input current levels between ±1.2 pA. Panel (a): Section of raw DUT readings from a calibration run, showing 3 ½ on-off cycles. The input current is stepped in units of 20 fA between each on-off cycle. Panel (b): section of raw reference current source readings from the same run as (a). Panels (a) and (b) share the same time axis, and vertical double arrows indicate the differences $\Delta I_M$ and $\Delta I_{REF}$. Panel (c): Inverse of the calibration factor plotted as a function of reference current.

Some example raw readings of $I_M$ and $I_{REF} = I_{IN}$ are shown in Figures 3(b) and 3(c) respectively. For the NPL reference current source, $I_{REF} = V_{REF}/R$, where $V_{REF}$ is the reading of a calibrated voltmeter (HP3458A) measuring the voltage drop across standard resistor R of nominal value 1 GΩ. Two things are worth noting from this raw data. First, the settling of the DUT readings is clearly apparent in the data of Fig. 3(b). Second, referring to the y-axis scales of the two plots, it is clear that $I_M$ is much noisier than $I_{REF}$. Almost all of the statistical uncertainty in the calibration factor is due to the DUT itself. Recalling from section 2.2 that the relative type B uncertainty in the reference current is less than $10^{-6}$, it is clear that the statistical uncertainty in the DUT readings is by far the dominant contribution to the overall uncertainty budget for C. As discussed in section 3, this is a common situation in small current calibrations. Repeating the calibration cycle 100 times (2000 seconds of measurement time) yields an averaged value for the calibration factor C. Figure 3(d) shows C measured over several days, exhibiting stability on the level of 10 ppm at each current level, but with a difference of around 100 ppm between the two current levels. During this calibration, the laboratory temperature was stable to ±1 °C and the sensitive pre-amplifier part of the DUT was not subjected to any mechanical shock. It has been found that even small mechanical shocks, for example due to connecting a cable to the pre-amplifier, can change the calibration factor by more than 50 ppm.

Finally, we compare the statistical uncertainty of this calibration with that predicted from the idealised noise model of section 3. The typical relative statistical uncertainty for one data point at 100 pA in fig. 3(d) is $5\times10^{-6}$, for 2000 s of measurement time, implying a relative uncertainty for a one-second measurement of $5\times10^{-6}\times\sqrt{2000} = 2.2\times10^{-4}$. This is plotted in figure 2 as open circles, along with the corresponding point for the 50 pA current level. These numbers are about a factor 10 higher than the uncertainty calculated for the idealised ammeter model. The discrepancy in this case is due to several factors: 1) the DUT is set to the 1 nA range, not the 100 pA range which would be optimal for a 100 pA input current. 2) The reference current source adds some extra noise. 3) The on-off cycle, necessary to remove offset currents, reduces the effective measurement time by a factor of 2 and a further reduction in the effective measurement time results from the need to reject data points affected by the DUT's settling behaviour [21].

### 4.2 Particle-counting electrometer

Here, we calibrate a specialised instrument, the Grimm model 5.705, designed to measure the concentration of particulates in air. The instrument measures very small currents, in the range 0-1 pA. The target uncertainty for the calibration factor is quite modest, around 1 %, but the focus of the calibration is to establish the linearity of the instrument over its full measurement range. To perform the calibration, the charge collection cup of the instrument was dis-assembled, and current was fed directly into the charge collection electrode. As in the previous study, we use the NPL reference current source. Since the focus is on measuring $C$ at many current levels, we only perform one on-off cycle for each current level. Figures 4(a) and (b) show sections of raw readings $I_M$ and $I_{REF}$, and Figure 4(c) shows $1/C$ as a function of the reference current. The DUT response exhibits non-linearity at the 5% level, and there is a discontinuity at ≈300 fA due to an internal range change in the DUT.

As in the previous case study, the measurement noise of the DUT dominates the overall uncertainty, and we can make a quantitative comparison between the relative statistical uncertainties in the measured data (open squares in Fig. 2) and the uncertainty predicted by the idealised ammeter model of section 3. As in the previous case study, the measured uncertainty is larger than that predicted from the ideal model. Test measurements of the base DUT noise when it was disconnected from the reference current source show that most of the increase in uncertainty is due to the thermal noise in the reference current source's 1 GΩ resistor.

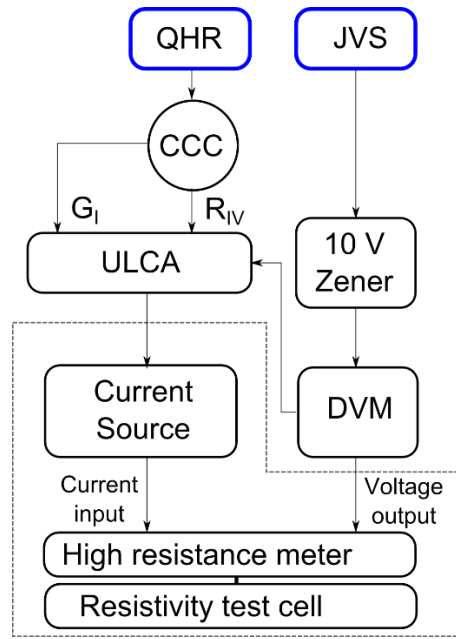

Figure 5. Schematic diagram of the traceability route used at NPL for measuring the resistivity of flat insulating sheet samples.

### 4.3 Resistivity of thin insulating sheets

In this case study, we describe a traceability path for a sheet resistivity measurement, illustrating the use of the ULCA as a stable reference current source. The resistivity of flat sheet samples is measured using a commercial high resistance meter, the HP 4392A, connected to a guarded test cell, the HP16008A. The HP 4392A measures resistance by applying a voltage and measuring a current, with the front panel ammeter dial functioning as an analogue computer to display the resistance in ohms as well as the current. The relative uncertainty in the resistivity is limited to around 5% by knowledge of the dimensions of the test cell, and we aim to make the resistance measurement with a relative uncertainty of around 1%. A diagram of the traceability path for this measurement is shown in figure 5. An ULCA, with gain traceable to the QHR primary resistance standard, is used to calibrate a commercial current source (in this case, a Keithley 6221) at a range of current levels from 10 pA to 10 nA, corresponding to resistances of 10 T$\Omega$ to 10 G$\Omega$ when biasing the sample with 100 V. The commercial current source is then used to generate a reference current for calibrating the current measurement function of the HP 4392A. The internal voltage source of the HP4392A is calibrated up to 100 V independently using a DVM (HP3458A) on the 100 V range, under the very reasonable assumption that the DVM has better than 1% linearity across the range. The traceability chain may seem complex and elaborate for a 1% measurement, but all the hardware outside the dotted box in Figure 4 has demonstrated annual drift of less than 10 ppm and therefore these calibrations need to be done very infrequently. The calibrations of the HP4392A, inside the dotted box, took less than 10 minutes to complete. It is worth pointing out that the commercial current source could be eliminated, and replaced by the ULCA operating in source mode. The reason this is not done at NPL is simply logistical: the single available ULCA unit is usually required for other measurements.

5. Conclusions

We have summarised recent developments in the field of small current metrology, and given an overview of best practice when calibrating small current ammeters.

- For small current calibrations, several traceability routes are available including recently developed ones based on the ULCA and electron pump. These offer advantages over the traditional capacitor ramp method, particularly where relative uncertainties smaller than 100 ppm are required.
- A plot of the Allan deviation of a time-series of measured current values can be used to quantitatively evaluate the stability of the instrumental offsets, and design the calibration cycle time to properly eliminate these offsets.
- A very simple noise model can be used to calculate the lower limit to the statistical uncertainty in a calibration (or indeed, any measurement) of a small current instrument.

Acknowledgements

This work was supported by the UK department for Business, Energy and Industrial Strategy and the EMPIR project 15SIB08. The European Metrology Programme for Innovation and Research (EMPIR) is co-financed by the Participating States and from the European Union's Horizon 2020 research and innovation programme. SPG would like to thank Dietmar Drung and Hansjoerg Scherer of the Physikalish-Technische Bundesanstalt for kindly loaning an ULCA, and Jonathan Williams for a careful reading of the paper.